\documentclass[notoc,a4paper,hyper]{JHEP3}

\usepackage[latin1]{inputenc}

\usepackage{amsmath}
\usepackage{amssymb}

\author{Domenico Orlando \\

Centre de Physique Théorique, École Polytechnique\footnote{Unité
    mixte du CNRS et de l'École Polytechnique, UMR 7644.} \\
  91128 Palaiseau, France

\bigskip

E-mail: \email{orlando@cpht.polytechnique.fr}
}

\title{\boldmath $\mathrm{AdS}_2 \times S^2$ as an exact heterotic string
  background\unboldmath}

\abstract{ An exact heterotic string theory on an $\mathrm{AdS}_2 \times S^2
  $ background is found as deformation of an $SL \left( 2, \mathbb{R}
  \right) \times SU \left( 2 \right)$ \textsc{wzw} model.\footnote{Based on a talk given at NATO
Advanced Study Institute and EC Summer School on String Theory: from Gauge
Interactions to Cosmology, Cargese, Corsica, France, 7 Jun - 19 Jun 2004.}  }  

\preprint{hep-th/0502213}

\begin{document}

\section{Intro}
\label{cha:intro}

Anti de Sitter in three dimensions and $S^3$ are among the most simple and
yet interesting string backgrounds. They are exact solutions to the string
equations beyond the supergravity approximation and, at the same time, are
simple to deal with although non-trivial thanks to the presence of
non-vanishing curvatures. For this reason they constitute an unique setting
in which to analyze AdS/\textsc{cft} correspondence, black-hole physics,
little-string theory.

String propagation in these backgrounds is described in terms of
\textsc{wzw} models for the $SL \left( 2, \mathbb{R}\right)$ and $SU \left(
  2 \right)$ groups, hence marginal deformations of such models allow to
study moduli space of the string vacua.  In particular well-known class of
marginal deformations for wzw models are those driven by left--right current
bilinears \cite{Chaudhuri:1989qb,Forste:2003km}. On the other hand $S^3$ and
$\mathrm{AdS}_3$ are embedded in larger structures so one can consider
marginal deformations where just one of the currents belongs to the $SU
\left( 2 \right)$ or $\mathrm{AdS}_3$ algebra, the other belonging to some
other $U\left( 1 \right)$ corresponding to an internal magnetic or electric
field.

This kind of deformation generates a continuous line of exact
\textsc{cft}'s. In this note we will show how with an appropriate choice for
the deforming current we obtain a boundary in moduli space and that this
boundary can be given a simple geometric interpretation
\cite{Israel:2004vv,Israel:2004cd} in terms of the $\mathrm{AdS}_2 \times
S^2 $ near-horizon geometry of the Bertotti-Robinson black hole
\cite{Bertotti:1959pf,Robinson:1959}.

\section{$SU \left( 2 \right)$ asymmetric deformation}
\label{cha:su-left-2}

In the $SU \left( 2 \right)$ case, there exists just one possible choice for
the deforming current the two other being related by inner automorphisms,
since the group has rank one, is compact and its Lie algebra simple. Take
the \textsc{wzw} model for $SU \left( 2 \right)$:
\begin{equation}
  S_{SU(2)_k} =\frac{1}{2\pi} \int \mathrm{d}^2 z \  \left\{\frac{k}{4}
    \left( \partial \alpha \bar \partial \alpha + \partial \beta \bar \partial \beta +
      \partial \gamma \bar \partial 
      \gamma + 2 \cos \beta \, \partial \alpha \bar \partial \gamma \right) +
    \sum_{a=1}^{3} \psi^a \bar \partial \psi^a\right\}
\end{equation}
where $\psi^a$ are the left-moving free fermions, superpartners of the
bosonic $SU(2)_k$ currents, and $(\alpha,\beta,\gamma)$ are the usual Euler
angles parameterizing the $SU(2)$ group manifold. The left-moving fermions
transform in the adjoint of $SU \left( 2 \right)$; there are no right-moving
superpartners but a right-moving current algebra of total charge $c= 16 $
can be realized in terms of right-moving free fermions. This means that we
can build a $\mathcal{N} = \left( 1, 0\right)$ world-sheet
supersymmetry-compatible deformation given by:
\begin{equation}
  \delta S_{\rm magnetic} = \frac{\sqrt{k k_G}H}{2\pi} \int {\rm
    d}^2 z \left(J^3 + \imath \psi^1 \psi^2\right) \bar J_G;  
\end{equation}
where $J^3 $ belongs to the $SU \left( 2 \right)$ algebra and $\bar J_g $ is
the current of the algebra at level $k_g$ realized by the right-moving free
fermions. An exact \textsc{cft} is obtained for any value of the deformation
parameter $H$.

\subsection{Geometry}
\label{sec:geometry}

These new backgrounds all present a constant dilaton, a magnetic field, a
\textsc{ns-ns} field proportional to the unperturbed one and a metric
retaining a residual $SU \left( 2 \right) \times U\left( 1 \right)$ isometry
\cite{Kiritsis:1995iu}. The most remarkable property is that the deformation
line in moduli space has a boundary corresponding to a critical value of the
deformation parameter $H^2=1/2$. At this point the $U \left( 1 \right)$
subgroup decompactifies and the resulting geometry is the left coset $SU
\left( 2 \right)/U\left( 1 \right) \sim S^2$ which is thus found to be an
exact \textsc{cft} background only supported by a magnetic field (the
dilaton remains constant and \textsc{ns} field vanishes). A geometrical
interpretation for this process can be given as follows: the initial $S^3 $
sphere is a Hopf fibration of an $S^1 $ fiber generated by the $J^3$ current
over an $S^2 $ base; the deformation only acts on the fiber, changing its
radius up to the point where this seems to vanish, actually marking the
trivialization of the fibration:
\begin{equation}
  S^3  \xrightarrow[H^2 \to H_{\text{max}}^2]{} \mathbb{R} \times
  S^2,
\end{equation}

If we turn our attention to the gauge field one can show that a quantization
of the magnetic charge is only compatible with levels of the affine algebras
such that $\frac{k}{k_G}=p^2 \ , \ \ p\in \mathbb{Z}$.  We will find the same
condition in terms of the partition function for the boundary deformation.

Although this construction has been implicitly carried on for first order in
$\alpha^\prime $ background fields, it is important to stress that the resulting
metric is nevertheless exact at all orders since the renormalization boils
down to the redefinition of the level $k $ that is simply shifted by the
dual Coxeter number (just as in the \textsc{wzw} case).

\subsection{Partition Function}
\label{sec:partition-function}

Consider the case of $k_g = 2$ (one right-moving $\mathbb{C}$ fermion). The
relevant components of the initial partition funciton are given by a
$SU(2)_{k-2}$-modular-invariance-compatible combination of $SU(2)_{k-2}$
supersymmetric characters and fermions from the gauge sector. For our
pourposes it is useful to further decompose the supersymmetric $SU(2)_{k}$
characters in terms of those of the $\mathcal{N}=2$ minimal models:
\begin{equation}
  \chi^j (\tau ) \ \vartheta {{a}\atopwithdelims[]{b}} (\tau , \nu ) = \sum_{m \in
    \mathbb{Z}_{2k}} \mathcal{C}^{j}_{m} {{a}\atopwithdelims[]{b}} \Theta_{m,k} \left( \tau,
    -\frac{2\nu}{k} \right) .
\end{equation}
The deformation acts as a boost on the left-lattice contribution of the
Cartan current of the supersymmetric $SU(2)_k$ and on the right current from
the gauge sector:
\begin{multline}
  \Theta_{m,k} \ \bar \vartheta {{h}\atopwithdelims[]{g}} = \sum_{n,\bar{n}} \mathrm{e}^{-\imath \pi
    g\left(\bar{n}+\frac{h}{2}\right)} q^{\frac{1}{2} \left(\sqrt{2k}
      n+\frac{m}{\sqrt{2k}}\right)^2}
  \bar{q}^{\frac{1}{2}\left(\bar{n}+\frac{h}{2}\right)^2}
  \\
  \longrightarrow \sum_{n,\bar{n}} \mathrm{e}^{-\imath\pi g\left(\bar{n}+\frac{h}{2}\right)}\ 
  q^{\frac{1}{2} \left[ \left(\sqrt{2k} n + \frac{m}{\sqrt{2k}} \right)\cosh
      x
      + \left( \bar n + \frac{h}{2} \right) \sinh x \right]^2} \\
  \times \bar{q}^{\frac{1}{2} \left[ \left( \bar n + \frac{h}{2} \right) \cosh x
      +\left(\sqrt{2k} n + \frac{m}{\sqrt{2k}} \right)\sinh x \right]^2} ,
\end{multline}
where the boost parameter $x$ is given by $ \cosh x = \frac{1}{1-2H^2}$.

Although an exact \textsc{cft} is obtained for any value of the deformation
parameter $H$ we will concentrate, as before, on the boundary value $H^2 =
1/2$. In this case the boost parameter diverges thus giving the following
constraints: $4(k+2)n+2m + 2 \sqrt{2k}\bar n + \sqrt{2k} h = 0$. Therefore,
the limit is well-defined only if the level of the supersymmetric $SU(2)_k$
satisfies the quantization condition $k = 2p^2 \ , \ \ p \in \mathbb{Z}$
\emph{i.e.} the charge quantization for the flux of the gauge field. Under
these constraints the $U(1)$ corresponding to the combination of charges
orthogonal our condition decouples and can be removed. In this way we end up
with the expression for the $S^2$ partition function:
\begin{equation}
  Z_{S^2} {{a ; h}\atopwithdelims[]{b ; g}} = \sum_{j,\bar \jmath} M^{j \bar \jmath}
  \sum_{N \in \mathbb{Z}_{2p}} \mathrm{e}^{\imath\pi g\left(N + \frac{h}{2}\right)}
  \ \mathcal{C}^{j}_{p(2N-h)} {{a}\atopwithdelims[]{b}} \ \bar{\chi}^{\bar
    \jmath}
\end{equation}
in agreement with the result found in~\cite{Berglund:1996dv} by using the
coset construction. The remaining charge $N$ labels the magnetic charge of
the state under consideration.

\section{$SL \left(2, R \right)$ deformation}
\label{sec:sl-left2-mathbbr}

The same construction as above can be repeated for the $SL \left( 2,
  \mathbb{R} \right)$ \textsc{wzw} model. In this case the moduli space is
somewhat richer for it is possible to realize three different asymmetric
deformations using the three generators of the group. These are not
equivalent ($SL \left( 2, \mathbb{R}\right)$ is not compact) and in fact
they lead to three physically different backgrounds.  The elliptic
deformation line, in example, contains the G\"odel universe
\cite{Israel:2003cx}, the parabolic deformation gives the superposition of
$\mathrm{AdS}_3$ and a gravitational plane wave.  Two of these deformation
lines present the same boundary effect as the $SU \left( 2 \right)$
deformation. In particular the elliptic deformation leads to the hyperbolic
space $\mathrm{H}_2 = SL \left( 2, \mathbb{R} \right)/U\left( 1 \right)$
supported by an immaginary magnetic field, \emph{ie} an exact but
non-unitary \textsc{cft}.  The hyperbolic deformation, on the other hand,
leads to $\mathrm{AdS}_2 = SL \left( 2, \mathbb{R} \right)/U\left( 1
\right)$ supported by an electric field. No charge quantization is present
in this case, because of the non-compact nature of the background.

In this latter case it is not yet possible to give the same construction for
the partition function as for the $SU \left( 2 \right)$ case since this
would require the decomposition of the initial partition function in a basis
of hyperbolic characters which is not a simple exercize. Nevertheless by
following the same procedure as before it is possible to evaluate the effect
of the deformation on the spectrum of primaries and hence give the resulting
$\mathrm{AdS}_2 $ background spectrum.

\section{AdS${}_2 \times S^2$}
\label{sec:mathrmads_2-times-s2}

The $S^2 $ and $\mathrm{AdS}_2 $ backgrounds can be combined so to give an
exact \textsc{cft} corresponding to the $\mathrm{AdS}_2 \times S^2 $ near-horizon
geometry of the \textsc{br} black-hole.

Let us now consider the complete heterotic string background which consists
of the $\mathrm{AdS}_2 \times S^2$ space--time times an $\mathcal{N}=2$ internal
conformal field theory $\mathcal{M}$, that we will assume to be of central
charge $\hat{c}=6$ and with integral $R$-charges.  The levels $k$ of $SU(2)$
and $\hat{k}$ of $SL(2,\mathbb{R})$ are such that the string background is
critical:
\begin{equation}
  \hat{c} = \frac{2(k-2)}{k} + \frac{2(\hat{k}+2)}{\hat{k}} =
  4 \implies k = \hat{k}.
\end{equation}
This translates into the equality of the radii of the corresponding $S^2$
and $\mathrm{AdS}_2$ factors, which is in turn necessary for supersymmetry.
Furthermore, the charge quantization condition for the two-sphere restricts
further the level to $k = 2p^2$, $p \in \mathbb{N}$.

The combined $\mathrm{AdS}_2 \times S^2 $ background can give new insights about
the physics of the \textsc{br} black hole in particular by analizing the
Schwinger-pair production in such background, or the study of the stability
and propagation of D-branes.

\begin{acknowledgments}
  Research partially supported by the EEC under the contracts
  HPRN-CT-2000-00131, HPRN-CT-2000-00148, MEXT-CT-2003-509661,
  MRTN-CT-2004-005104 and MRTN-CT-2004-503369.
\end{acknowledgments}

\bibliography{Biblia}

\end{document}